\documentclass[prl,preprint,showpacs,preprintnumbers,superscriptaddress,amsmath,amssymb]{revtex4}

\usepackage[dvips]{color}
\usepackage{graphicx}
\usepackage{dcolumn}
\usepackage{bm}
\usepackage{latexsym}

\preprint{}

\begin{document}

\title{Anisotropy of the in-plane resistivity of underdoped Ba(Fe$_{1-x}$Co$_x$)$_2$As$_2$ superconductors induced by impurity scattering in the antiferromagnetic orthorhombic phase}

\author{S. Ishida}
 \affiliation{Department of Physics, University of Tokyo, Tokyo 113-0033, Japan}
 \affiliation{National Institute of Advanced Industrial Science and Technology, Tsukuba 305-8568, Japan}
 \affiliation{JST, Transformative Research-Project on Iron Pnictides, Tokyo 102-0075, Japan}
 
\author{M. Nakajima}
 \affiliation{Department of Physics, University of Tokyo, Tokyo 113-0033, Japan}
 \affiliation{National Institute of Advanced Industrial Science and Technology, Tsukuba 305-8568, Japan}
 \affiliation{JST, Transformative Research-Project on Iron Pnictides, Tokyo 102-0075, Japan}
 
\author{T. Liang}
 \affiliation{Department of Physics, University of Tokyo, Tokyo 113-0033, Japan}
 \affiliation{National Institute of Advanced Industrial Science and Technology, Tsukuba 305-8568, Japan}
 \affiliation{JST, Transformative Research-Project on Iron Pnictides, Tokyo 102-0075, Japan}

\author{K. Kihou}
 \affiliation{National Institute of Advanced Industrial Science and Technology, Tsukuba 305-8568, Japan}
 \affiliation{JST, Transformative Research-Project on Iron Pnictides, Tokyo 102-0075, Japan}
 
\author{C. H. Lee}
 \affiliation{National Institute of Advanced Industrial Science and Technology, Tsukuba 305-8568, Japan}
 \affiliation{JST, Transformative Research-Project on Iron Pnictides, Tokyo 102-0075, Japan}

\author{A. Iyo}
 \affiliation{National Institute of Advanced Industrial Science and Technology, Tsukuba 305-8568, Japan}
 \affiliation{JST, Transformative Research-Project on Iron Pnictides, Tokyo 102-0075, Japan}
 
\author{H. Eisaki}
 \affiliation{National Institute of Advanced Industrial Science and Technology, Tsukuba 305-8568, Japan}
 \affiliation{JST, Transformative Research-Project on Iron Pnictides, Tokyo 102-0075, Japan}
 
\author{T. Kakeshita}
 \affiliation{Department of Physics, University of Tokyo, Tokyo 113-0033, Japan}
 \affiliation{JST, Transformative Research-Project on Iron Pnictides, Tokyo 102-0075, Japan}

\author{Y. Tomioka}
 \affiliation{National Institute of Advanced Industrial Science and Technology, Tsukuba 305-8568, Japan}
 \affiliation{JST, Transformative Research-Project on Iron Pnictides, Tokyo 102-0075, Japan}
 
\author{T. Ito}
 \affiliation{National Institute of Advanced Industrial Science and Technology, Tsukuba 305-8568, Japan}
 \affiliation{JST, Transformative Research-Project on Iron Pnictides, Tokyo 102-0075, Japan}
 
\author{S. Uchida}
 \affiliation{Department of Physics, University of Tokyo, Tokyo 113-0033, Japan}
 \affiliation{JST, Transformative Research-Project on Iron Pnictides, Tokyo 102-0075, Japan}

{\color{red}}

\date{\today}

\begin{abstract}
\indent We investigated the in-plane resistivity anisotropy for underdoped Ba(Fe$_{1-x}$Co$_x$)$_2$As$_2$ single crystals with improved quality. We demonstrate that the anisotropy in resistivity in the magnetostructural ordered phase arises from the anisotropy in the residual component which increases in proportion to the Co concentration $x$. This gives evidence that the anisotropy originates from the impurity scattering by Co atoms substituted for the Fe sites, rather than so far proposed mechanism such as the anisotropy of Fermi velocities of reconstructed Fermi surface pockets. As doping proceeds to the paramagnetic-tetragonal phase, a Co impurity transforms to a weak and isotropic scattering center. 
\end{abstract}

\maketitle


\indent Recent resistivity measurements performed on detwinned single crystals of Ba(Fe$_{1-x}TM_x$)$_2$As$_2$ ($TM$ = Co, Ni and Cu)~\cite{Chu,Kuo} revealed an anomalous in-plane resistivity anisotropy in the magnetostructural ordered phase (antiferromagnetic-orthorhombic (AFO) phase) - the resistivity is lower along antiferromagnetic and longer $a$ axis than along ferromagnetic and shorter $b$ axis. In addition, the magnitude of in-plane resistivity anisotropy is even enhanced by $TM$ substitution despite that the antiferromagnetic spin order weakens and the lattice orthorhombicity diminishes~\cite{Prozorov}. These results have been ascribed to the anisotropy of the reconstructed Fermi surface (FS) pockets in the AFO ordered state and anisotropic (stripe-like) spin structure~\cite{Kuo,Fisher}. However, FS pockets observed by angle-resolved photoemission spectroscopy (ARPES)~\cite{Yi1,Yi2,Y.K.Kim} and quantum oscillations~\cite{Analytis,Terashima} seem not sufficiently anisotropic to explain the observed large in-plane resistivity anisotropy. Furthermore, even the ellipsoidal FS pockets would result in almost isotropic conductivity when summed over the momentum space. It seems that different mechanisms are required to explain the large in-plane resistivity anisotropy and its doping dependence. By contrast, the in-plane resistivity anisotropy of hole-doped Ba$_{1-x}$K$_x$Fe$_2$As$_2$ is very small in sharp contrast to the electron doped cases~\cite{J.J.Ying}. It is suggested that the different magnetic states between the electron- and hole-doped materials may be responsible for the difference in the resistivity anisotropy. Nevertheless, this raises a question why only a slight electron- or hole-doping can cause such a dramatic difference in the electronic anisotropy. 

\indent Another notable feature is a finite in-plane resistivity anisotropy above the structural transition temperature $T_s$~\cite{Chu}. Maybe or may not be corresponding to this result, the orbital polarization - an energy splitting or a slight difference in electron occupancy between d$_{xz}$- and d$_{yz}$-orbital derived bands - is observed by ARPES above $T_s$~\cite{Yi1}. These results are interpreted as evidences for the nematic electronic state. However, the application of uniaxial pressure for detwinning breaks the lattice four fold symmetry, and so may stabilize the AFO order to higher temperatures. Recent X-ray diffraction~\cite{Blomberg2} and neutron scattering~\cite{Dhital} measured on detwinned BaFe$_2$As$_2$ single crystals have suggested that the applied pressure leads to an increase in the onset temperature $T_s$ (or the magnetic ordering temperature $T_{\rm N}$) to 150~K under the pressure of 0.7~MPa~\cite{Dhital}. If this is the case, the anisotropy above $T_s$ (of free-standing crystal) might not be an intrinsic property of this system. 

\indent In this letter, we present the results of in-plane resistivity measurements on detwinned Ba(Fe$_{1-x}$Co$_x$)$_2$As$_2$ single crystals with quality much improved by annealing. We previously reported that the in-plane resistivity is almost isotropic in the undoped BaFe$_2$As$_2$ after the crystal quality is improved by annealing~\cite{Ishida}. Below it will be shown that the magnitude of residual resistivity both in $a$- and $b$-direction as well as their anisotropy remarkably increase in proportion to the doped Co content $x$ in the AFO phase. This result gives evidence that the in-plane resistivity anisotropy originates from Co impurity scattering. The diminished in-plane resistivity anisotropy in hole-doped Ba$_{1-x}$K$_x$Fe$_2$As$_2$ is understood as a consequence of very weak carrier scattering of doped K atoms since the dopant site is away from the Fe plane. \\



\begin{figure*}[t!]
\includegraphics[width=\columnwidth,clip]{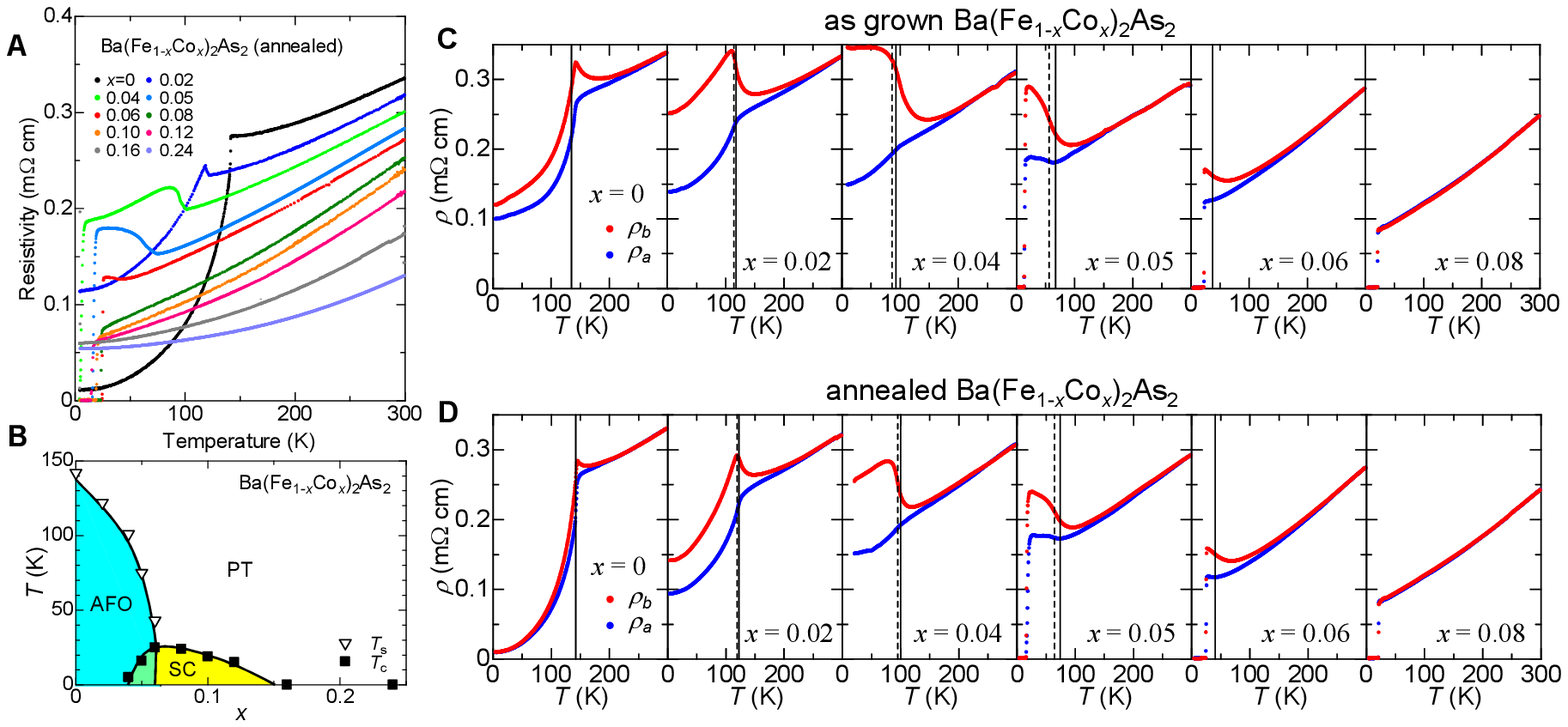}
\caption{\label{fig1} (Color online) (a) Temperature dependence of the in-plane resistivity for twinned annealed Ba(Fe$_{1-x}$Co$_x$)$_2$As$_2$ crystals with various compositions. (b) Phase diagram of Ba(Fe$_{1-x}$Co$_x$)$_2$As$_2$ determined from the results of in-plane resistivity measurement shown in (a). Triangles and squares indicate $T_s$ and $T_c$, respectively. (c) Temperature dependence of the in-plane resistivity $\rho_{a}$ (blue) and $\rho_{b}$ (red) measured on detwinned as-grown Ba(Fe$_{1-x}$Co$_x$)$_2$As$_2$ crystals. Values of $x$ are labeled in each panel. Vertical lines indicate $T_s$ (solid line) and $T_{\rm N}$ (dashed line) determined from $d\rho$($T$)/$dT$ under unstressed/twinned conditions, respectively. (d) The same plots for the annealed crystals.}
\end{figure*}

\indent Single crystals of Ba(Fe$_{1-x}$Co$_x$)$_2$As$_2$ were grown by the self-flux method as described elsewhere~\cite{Nakajima1}. The Co compositions of the samples were determined by inductively coupled plasma (ICP) analysis. In Fig. \ref{fig1} we show the phase diagram of Ba(Fe$_{1-x}$Co$_x$)$_2$As$_2$ obtained from the in-plane resistivity measurement on twinned single crystals. Typical dimension of the crystal is 1.5$\times$1.5~mm$^2$ in the $ab$-plane area and 0.5~mm in thickness along the $c$ axis. The crystals were sealed in an evacuated quartz tube together with BaAs powders and annealed for several days at 800~$^{\mathrm{o}}$C. This annealing was found to remarkably improve transport properties in the ordered phase of BaFe$_2$As$_2$~\cite{Ishida,Rotundu}, which indicates that the as-grown crystals contain appreciable amount of defects/impurities, and hence the observation of the intrinsic charge transport in this system might be disturbed. The magnitude and temperature ($T$) dependence of resistivity look different from those reported by Chu $et~al.$~\cite{Chu}, particularly in the AFO phase. For the measurement of in-plane resistivity anisotropy, Ba(Fe$_{1-x}$Co$_x$)$_2$As$_2$ crystals were set in an uniaxial pressure cell and detwinned by applying compressive pressure along the tetragonal (110) direction~\cite{Liang}. The resistivities along the $a$ and $b$ axis were measured simultaneously using the Montgomery method~\cite{Montgomery} without releasing pressure. The measurements were performed in a Quantum Design physical property measurement system (PPMS). \\


\indent Figure \ref{fig1} shows the $T$ dependence of the in-plane resistivity measured on detwinned Ba(Fe$_{1-x}$Co$_x$)$_2$As$_2$ crystals ($x$ = 0, 0.02, 0.04, 0.05, 0.06 and 0.08) (c) before and (d) after annealing. The data in blue and red color correspond to the resistivity along $a$ axis ($\rho_a$) and $b$ axis ($\rho_b$), respectively. Vertical lines indicate the structural transition temperature $T_{\rm s}$ (solid line) and magnetic transition temperature $T_{\rm N}$ (dashed line) determined from the $T$ dependence of resistivity measured on twinned (free standing) crystals following the method in Ref.~\cite{Chu2}. At $T_s$ the in-plane resistivity ($\rho_{ab}$) suddenly drops for $x$ = 0 or rapidly increases for $x$ $\neq$ 0, and $T_{\rm N}$ is determined at a temperature where $\rho_{ab}$ shows a cusp ($x$ = 0.02) or (-d$\rho$/d$T$) shows a peak. $T_s$ = $T_{\rm N}$ for $x$ = 0, but a difference between $T_s$ and $T_{\rm N}$ ($T_s$ $>$ $T_{\rm N}$) gets bigger as Co doping proceeds, but both temperatures are ill defined for $x$ $>$ 0.05. In view of what happen in $\rho_{ab}$ at $T_s$ and $T_{\rm N}$, it follows that for detwinned crystals $\rho_a$ changes its slope at $T_s$ and $\rho_b$ shows a cusp or its derivative (-$d\rho_b$/$dT$) shows a peak at $T_{\rm N}$. Then, the result of anisotropic resistivity for the as-grown crystal with $x$ = 0 indicates that $T_s$ ($T_{\rm N}$) shifts to higher temperature by the applied uniaxial pressure, consistent with the X-ray diffraction~\cite{Blomberg2}  and the neutron scattering~\cite{Dhital} measurements. The pressure-induced shift becomes smaller with increasing Co content. On the other hand, for annealed crystals, even for $x$ = 0, there appears no appreciable shift of $T_s$ and $T_{\rm N}$ by detwinning. 

\indent The resistivity anisotropy is not large but finite for the as-grown BaFe$_2$As$_2$ ($x$ = 0). The results for as grown crystals (Fig. \ref{fig1}(c)) reproduce those in the previous reports~\cite{Chu}, except for that the magnitude of resistivity in the paramagnetic-tetragonal (PT) phase exhibits a systematic decrease with doping in the present work. The results for annealed Ba(Fe$_{1-x}$Co$_x$)$_2$As$_2$ crystals (Fig. \ref{fig1}(d)) are different from as-grown ones, especially for low-doping regime ($x$ = 0 and 0.02). The resistivity anisotropy and the magnitude of residual resistivity are remarkably decreased after annealing. In particular, the anisotropy almost vanishes at low temperatures for $x$ = 0. However, when Co is doped, both rapidly increase up to $x$ = 0.04 just before superconductivity emerges. The temperature range of the anisotropy above $T_s$ also widens by Co doping. With further doping Co the anisotropy turns to decrease. For $x$ = 0.05 and 0.06, annealing effect is weak with both residual resistivity and the in-plane resistivity anisotropy only slightly decreased after annealing. No anisotropy is observed for optimally doped ($x$ = 0.08) compound, and there is no visible effect of annealing. Notably, $\rho_a$ and $\rho_b$ for annealed crystals show similar $T$ dependence in the AFO phase. This suggests that the $T$-dependent component of resistivity is nearly isotropic, and dominant contribution to the resistivity anisotropy stems from the anisotropic residual resistivity. 

\indent As regards $T$ dependence of resistivity, in the case of as-grown $x$ = 0.04 sample, $\rho_a$ and $\rho_b$ show different $T$ dependence in the AFO phase; $\rho_a$ is metallic whereas $\rho_b$ is nearly $T$-independent. The reason for this difference is probably due to the inhomogeneity of Co concentration and/or the crystal defects. As a consequence of such disorder, the transition broadens and the AFO order may not develop over the entire sample even at low temperatures. At higher doping levels, $x$ $\geq$ 0.05, neither $\rho_a$ nor $\rho_b$ exhibit clear changes any more at the transition temperature. This is because the AFO order weakens so that the compounds become relatively insensitive to disorder, dopant Co atom, and lattice defects, which exist before annealing. 

\indent For crystals with $x$ = 0, 0.02, and 0.04, the PT-AFO transition sharpens after annealing, and the temperature range of the anisotropy above $T_s$ is narrowed. It is expected for $x$ =0 (and maybe for $x$ = 0.02) that in the clean limit the anisotropy above $T_s$ vanishes as in the case of SrFe$_2$As$_2$ and CaFe$_2$As$_2$ where the in-plane resistivity anisotropy above $T_s$ is negligibly small (the phase transition is essentially of first order)~\cite{Blomberg1}. 

\indent These results suggest that the resistivity anisotropy persisting above $T_s$ is due to the spacial inhomogeneity of the Co impurity concentration and/or lattice defects rather than due to the existence of the nematicity. If the nematicity were to exist, by annealing the magnitude of  resistivity anisotropy above $T_s$ as well as the temperature where $\rho_a$ and $\rho_b$ start to split ($T^*$) would be increased, since $T_s$ increases after annealing which indicates that the anisotropic electronic states becomes more stable. This is in contradiction to the experimental result that the annealing decreases the overall resistivity anisotropy as well as $T^*$. Thus, the observed resistivity anisotropy above $T_s$ is probably extrinsic in origin as in the case of the anisotropy in the AFO phase. 

\indent As we previously reported, in the AFO ordered phase of the annealed BaFe$_2$As$_2$, the residual resistivity remarkably decreases and the in-plane resistivity anisotropy becomes vanishingly small at low temperatures~\cite{Ishida}. From the results of magneto-transport measurements we concluded that some carriers in the AFO phase have high and isotropic mobility and mask the intrinsically anisotropic charge dynamics revealed by the measurement of optical conductivity~\cite{Nakajima2}. These high-mobility carriers would arise from Dirac-cone shaped energy bands, inherent to the unique topological nature of the band structure of BaFe$_2$As$_2$ in the AFO phase~\cite{Ran,Morinari}.

\indent It is unlikely that only 2~$\%$ Co substitution wipes out the Dirac Fermi pocket and unmasks the intrinsic anisotropy and/or make the Fermi surface or the effective mass of carriers dramatically anisotropic~\cite{Liu}. To the contrary, Co doping weakens the AFO order which involves anisotropic spin alignment and uniaxial lattice distortions, both being possible sources of anisotropy. If the carriers maintain nearly isotropic effective mass, then, a question is what makes the in-plane resistivity anisotropic? In order to solve this problem, we analyze the data of annealed crystals in more details, in which genuine Co doping effect is expected to show up in the charge transport. 

\begin{figure}[t]
\includegraphics[width=0.55\columnwidth,clip]{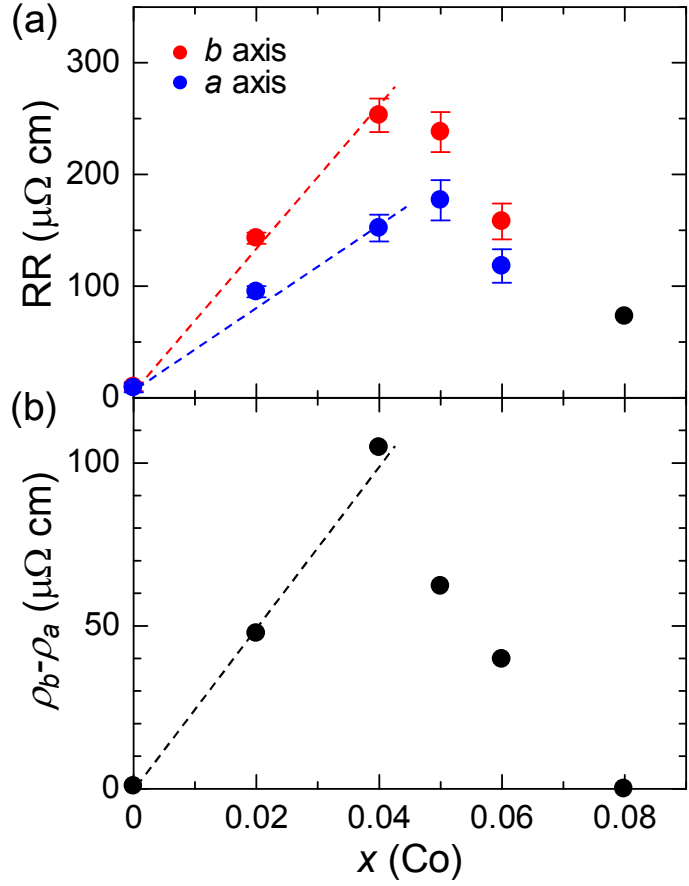}
\caption{\label{fig2} (Color online) (a) Residual resistivity and (b) difference of in-plane resistivity ($\rho_b$-$\rho_a$) at low temperatures plotted against the Co content for annealed Ba(Fe$_{1-x}$Co$_x$)$_2$As$_2$ crystals. Dashed lines are guides to the eye.}
\end{figure}

\indent Figure \ref{fig2}(a) shows the doping dependence of the residual resistivity $\rho_0$ in the $a$- and $b$-axis direction. $\rho_0$ values of BaFe$_2$As$_2$ in both $a$ and $b$ directions decreases to $\sim$10~$\mu\Omega$cm after annealing, by an order lower than that of as-grown crystal. Upon Co doping a very large residual resistivity is produced in both directions. As guided by dashed lines, the increase in the residual resistivity is nearly proportional to the Co content in the ``pure'' AFO regime ($x$ $\leq$ 0.04). In particular, its increase in the $b$ direction is remarkable. Consequently, not only the residual resistivity, but also $\rho_b$ - $\rho_a$ is proportional to the Co concentration up to $x$ = 0.04, as plotted in Fig. \ref{fig2}(b). The $\rho_0$ value along the $b$ axis well exceeds 100~$\mu\Omega$cm  for 2~$\%$ substitution of Co atoms, which is comparable with $\rho_0$ induced by Zn doping into underdoped cuprates~\cite{Fukuzumi}. 

\indent To summarize present results, the magnitude of both the residual resistivity component and the resistivity anisotropy increase linearly with the doped Co content. Below, we discuss the microscopic origin of the in-plane resistivity anisotropy. 

\begin{figure}[t]
\includegraphics[width=0.65\columnwidth,clip]{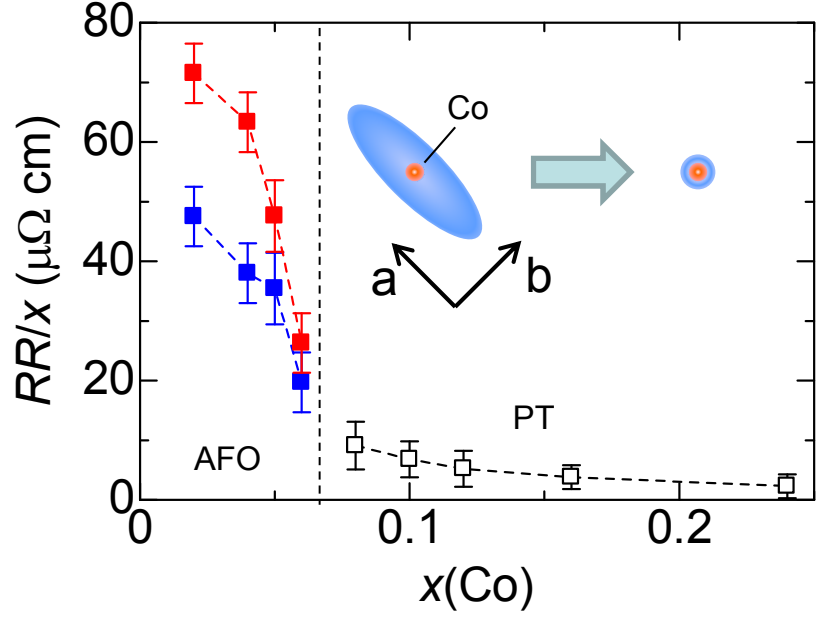}
\caption{\label{fig3} (Color online) Doping dependence of residual resistivity ($RR$/$x$ normalized by Co concentration $x$ in $\%$) in the AFO phase (along $a$ axis (blue squares) and $b$ axis (red squares)) and in the PT phase (black open squares) for annealed Ba(Fe$_{1-x}$Co$_x$)$_2$As$_2$ crystals. Schematic view of a Co impurity state working as a scattering center is shown in the inset.}
\end{figure}

\indent From the results we reconfirm that the in-plane resistivity anisotropy originates from the anisotropy in the residual resistivity component. This implies that a doped Co atom works as an anisotropic scattering center with scattering cross section larger for carriers propagating along the $b$-axis (see a cartoon in the inset to Fig. \ref{fig3}). According to the theoretical calculation~\cite{Nakamura}, in the PT phase Co substituted for Fe introduces a weak and isotropic impurity potential of $\sim$~-~0.5~eV to the d-electrons, much smaller than the band width of the d-bands. However, probably reflecting the underlying anisotropic electronic environment of the AFO phase, a Co atom anisotropically polarizes its surrounding and thereby scatters carriers more strongly in the $b$-direction. This is similar to the case of Zn in the cuprates. Zn antiferromagnetically polarizes its surrounding Cu spins~\cite{Julien} and acts as a strong scatterer in the unitarity limit. It is theoretically speculated that the induced anisotropic polarization may be associated with formation of local orbital order around a Co atom, not spin order~\cite{Kontani}. 

\indent The present conclusion is confirmed also by the optical measurement on detwinned Ba(Fe$_{1-x}$Co$_x$)$_2$As$_2$ crystals from the same batch~\cite{Nakajima3} which finds that the Co doping makes the width of a Drude component broader and anisotropic in the AFO phase. The Drude width is significantly larger along the $b$ axis than the $a$ axis, in quantitative agreement with the resistivity anisotropy in the present measurement. Moreover, the recent scanning-tunneling-spectroscopy (STS) measurement has been successful in observing the formation of $a$-axis aligned electronic dimers surrounding each Co in the AFO state of Ca(Fe$_{1-x}$Co$_x$)$_2$As$_2$~\cite{Allan}, in agreement with our speculation of the anisotropic Co impurity state. 

\indent Certainly, formation of an $a$-axis elongated anisotropic Co impurity state is responsible for the jump-up of $\rho_b$ near $T_s$. In the light of the present result, the ``precursory" increase of $\rho_b$ from above $T_s$ likely originates from the enhanced $b$-axis impurity scattering due to the presence of short-range AFO order. Short-range AFO order might be formed at $T$ $>$ $T_s$ in the region where Co density is relatively low, and the AFO order might not develop in the region where Co density is high until $T$ is lowered well below $T_s$. The gradual increase of  $\rho_b$ with decrease of $T$ across $T_s$ for $x$ $\geq$ 0.04 would be due to this disorder effect. 

\indent When the doping exceeds $x$ = 0.04 where superconductivity appears and coexists with the AFO order, both in-plane resistivity anisotropy and residual resistivity rapidly decrease. This suggests that at these doping levels the AFO order gets too weak for Co impurity to polarize its surroundings. In Fig. \ref{fig3} we plot the doping dependence of residual resistivity ($RR$/$x$ normalized by the Co concentration $x$ in $\%$), a measure of scattering strength of individual Co impurity. In the AFO phase $RR$/$x$ is $\sim$~60~$\mu\Omega$cm/$\%$ ($x$ = 0.02), and becomes by an order of magnitude smaller in the PT phase above $x$ = 0.08. Considering the Drude formula $\rho_0$ = $\frac{m^*}{n_{\rm D}e^2} \cdot \frac{1}{\tau}$ with the effective mass $m^*$ and the density $n_{\rm D}$ of relavant carriers and the scattering time $\tau$, one might consider that this is due to the increase of carrier density. However, from the result of optical measurement~\cite{Nakajima1}, the spectral weight $n_{\rm D}$/$m^*$ of the coherent Drude component, which dominates the low-energy conductivity spectrum, is unchanged across the AFO-PT transition and increase by only 60$\%$ on going from $x$ = 0 to $x$ = 0.08. Therefore, the decrease of $RR$/$x$ is ascribed to a radical weakening of the scattering strength rate 1/$\tau$, and Co transforms to a weak and isotropic scattering center in the highly doped region. When an impurity potential is weak and forward scattering dominates, the impurity is not destructive even for the $s_{\pm}$- or $d$-wave superconductivity with sign changing gap~\cite{Zhu}. 

\indent The present impurity-induced-anisotropy scenario explains the tiny in-plane resistivity anisotropy observed for Ba$_{1-x}$K$_x$Fe$_2$As$_2$. Since K is substituted for the Ba site located away from the Fe plane, carrier scattering from the K impurities is so weak that the doped K atom does not polarize its surrounding anisotropically. Weak scattering potential of K is evidenced by the magnitude of the residual resistivity of Ba$_{1-x}$K$_x$Fe$_2$As$_2$ in the AFO phase which is by an order of magnitude smaller than that of Ba(Fe$_{1-x}$Co$_x$)$_2$As$_2$. Recently, it has been reported that the electron-doped Ca$_{1-x}R_x$Fe$_2$As$_2$ ($R$ = rare earth elements) shows large in-plane resistivity anisotropy even though the dopant $R$ site is away from the Fe plane~\cite{J.J.Ying2}. However, the samples of Ca$_{1-x}R_x$Fe$_2$As$_2$ show an anomalously large residual resistivity which does not increase with $x$. Origin of this large residual resistivity is likely lattice defects and/or impurities introduced nearby the Fe planes as in the case of the as-grown BaFe$_2$As$_2$. Thus, whether the system is doped with electrons or holes does not matter in producing the in-plane resistivity anisotropy. It is induced by strong disorder. \\

\indent In summary, we investigated the origin of the in-plane resistivity anisotropy of Ba(Fe$_{1-x}$Co$_x$)$_2$As$_2$ using single crystals with much improved quality. It is found that magnitudes of anisotropy in the residual resistivity is proportional to the Co content in the AFO phase. This result evidences that the in-plane resistivity anisotropy is extrinsic in origin and arises from the elastic impurity scattering by doped Co impurities, not from the anisotropic Fermi surface nor inelastic scattering process involving anisotropic spin or charge/orbital excitations in the AFO ordered phase. Substituted Co atom would polarize its environment with inherent electronic anisotropy in the AFO phase, thereby acquiring an anisotropic scattering cross-section. In the PT phase, the Co impurity totally changes its nature, and works as a weak and isotropic scattering center. \\

\indent SI and MN thank to the Japan Society for the Promotion of Science (JSPS) for the financial support. This work was supported by Transformative Research-Project on Iron Pnictides (TRIP) from the Japan Science and Technology Agency, and by the Japan-China-Korea A3 Foresight Program from JSPS, and a Grant-in-Aid of Scientific Research from the Ministry of Education, Culture, Sports, Science, and Technology in Japan.

%
%

%
\newpage
%
%

\end{document}